\documentclass[11pt]{article}
\usepackage{moriond,epsfig}
\usepackage{amsmath,amssymb}

\def\cH{{\cal H}}
\def\cG{{\cal G}}
\def\cO{{\cal O}}
\def\and{\quad\text{and}\quad}
\begin{document}
\vspace*{4cm}
\title{THRESHOLD AND TRANSVERSE-MOMENTUM RESUMMATIONS \\ FOR GAUGINO-PAIR HADROPRODUCTION}
\author{Jonathan Debove}
\address{Laboratoire de Physique Subatomique et de Cosmologie, Universit\'e Joseph Fourier/CNRS-IN2P3/INPG, 53 Avenue des Martyrs, F-38026 Grenoble, France}
\maketitle
\abstracts{
We present precision calculations of the invariant-mass and transverse-momentum distributions of gaugino pairs produced at hadron colliders.
We implement the threshold and transverse-momentum resummation formalisms at next-to-leading order accuracy and match the obtained result to the perturbative prediction at $\cO(\alpha_s)$.
We compare the various resummed cross sections with the perturbative results and Monte Carlo predictions.
The theoretical uncertainties coming from renormalisation and factorisation scale variations are also discussed.
}

\section{Introduction}
The Minimal Supersymmetric Standard Model (MSSM) continues to be a very attractive extension of the Standard Model of particle physics.\cite{Nilles:1983ge,Haber:1984rc}
If $R$-parity is conserved, it provides a convincing candidate for the dark matter observed in the universe.
In the MSSM, this is generally the lightest neutralino, one of the fermionic partners of the electroweak gauge and Higgs bosons which mix to give four neutral (charged) mass eigenstates, namely the neutralinos (charginos). 
Light enough to be produced at current hadron colliders, these particles have been studied extensively, and the cross sections for the production of gaugino-pairs are known at leading order \cite{Barger:1983wc,Dawson:1983fw,Debove:2008nr} and next-to-leading order of perturbative QCD.\cite{Beenakker:1999xh}

In this paper, we extend this last work and resum the soft-gluon contributions which appear in both the invariant-mass ($M$) and transverse-momentum ($p_T$) distributions of the gaugino pairs.
Indeed, starting at $\cO(\alpha_s)$, logarithmic terms of the forms $[\alpha_s\ln(1-z)/(1-z)]_+$ and $\alpha_s\ln(M^2/p_T^2)/p_T^2$ appear where $z=M^2/s$ and $\sqrt{s}$ is the partonic centre-of-mass energy.
These terms become large close to the production threshold and in the small-$p_T$ region, thus possibly spoiling the convergence of the usual perturbative calculations.
In order to gain control over these terms, they must be resummed to all order in $\alpha_s$.

\section{Resummation formalisms}
The methods to systematically perform all-order resummation of classes of enhanced logarithms are well known and are generally performed in conjugate spaces.\cite{Sterman:1986aj,Catani:1989ne,Collins:1984kg,Bozzi:2005wk}
Working with the Mellin moments of the differential partonic cross sections, the resummed contributions take the factorised forms
\begin{align} M^2 \frac{d \sigma^{\rm(res)}}{d M^2}(N) &= H(M) \exp[G(N)] \and
    \\ \frac{M^2 d \sigma^{\rm(res)}}{d M^2 d p_T^2}(N) &= \int \frac{d^2 b}{4 \pi} e^{i \bf b \cdot p_T} \cH(M, N) \exp[\cG(N, Mb)]
\end{align}
for $M$- and $p_T$-distributions respectively.
The Mellin variable $N$ and the impact parameter $b$ are the conjugate variables of $z$ and $p_T$.
In these factorised expressions, all the potentially large logarithmic terms are embodied and resummed in the exponential form factors $\exp[G]$ and $\exp[\cG]$.
The process-dependent functions $H$ and $\cH$ do not depend on the relevant conjugate variables and can thus be computed perturbatively.
The general expressions for the functions $H$, $G$ and $\cH$, $\cG$ can be found in Refs.~\cite{Vogt:2000ci,Debove:2010kf} and Refs.~\cite{Bozzi:2005wk,Debove:2009ia}, respectively.

Once the large logarithms have been resummed in $N$- and $b$-space, we have to switch back to the physical spaces in order to achieve a phenomenological study.
Special attention has to be paid to the singularities in the exponents $G$ and $\cG$.
They are related to the presence of the Landau pole in the perturbative running of $\alpha_s$, and a prescription is needed.
In our numerical study, we follow Refs.~\cite{Catani:1996yz,Laenen:2000de} and deform the integration contour in the complex $N$- and $b$-planes.
Finally, in order to conserve the full information contained in the fixed-order calculation, the $\cO(\alpha_s)$ and the resummed 
calculations are matched by subtracting from their sum the expansion of the resummed cross section truncated at $\cO(\alpha_s)$
\begin{equation}
    d\sigma = d\sigma^{\rm(f.o)} + d\sigma^{\rm(res)} - d\sigma^{\rm(exp)},
\end{equation}
thus avoiding possible double counting of the logarithmically enhanced contributions.

\section{Numerical results}
We now present numerical results for the production of gaugino pairs at the Tevatron and at the LHC.
Unless stated otherwise, the parton densities are evaluated in the most recent parametrisation of the CTEQ collaboration {\tt CTEQ6.6M} \cite{Nadolsky:2008zw} with $\alpha_s$ evaluated at two-loop accuracy.
In the following, we choose the minimal supergravity benchmark point SPS1a' \cite{AguilarSaavedra:2005pw} and obtain the weak-scale supersymmetric parameters through the computer code {\tt SuSpect2.3}.\cite{Djouadi:2002ze}
The obtained gaugino masses are $m_{\tilde{\chi}_1^0} = 61$ GeV and $m_{\tilde{\chi}_2^0} = m_{\tilde{\chi}_1^\pm} = 183$ GeV.

\begin{figure}[t]
    \begin{minipage}[t]{.48\columnwidth}
	\epsfig{figure=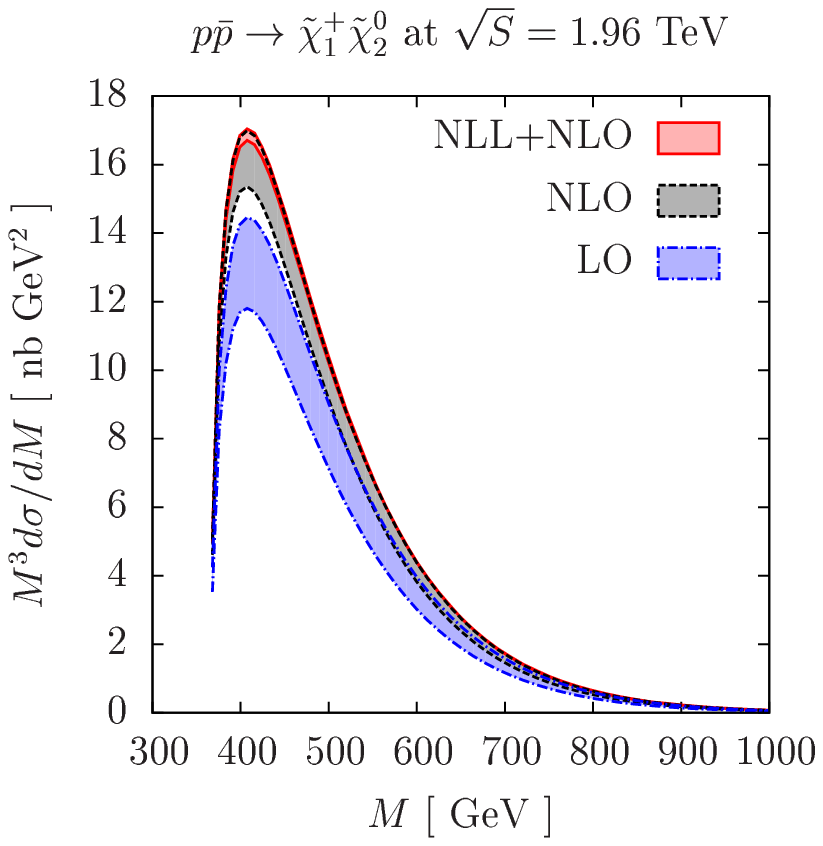,width=.9\textwidth}
	\caption{Invariant-mass distribution of chargino-neutralino pairs at the Tevatron.
	The LO calculation (dot-dashed) is compared to the resummation prediction (full) matched to the NLO calculation (dashed).
	The scale uncertainties are shown as shaded bands. \hfill
	\label{fig:1}}
    \end{minipage}\hfill
    \begin{minipage}[t]{.48\columnwidth}
	\epsfig{figure=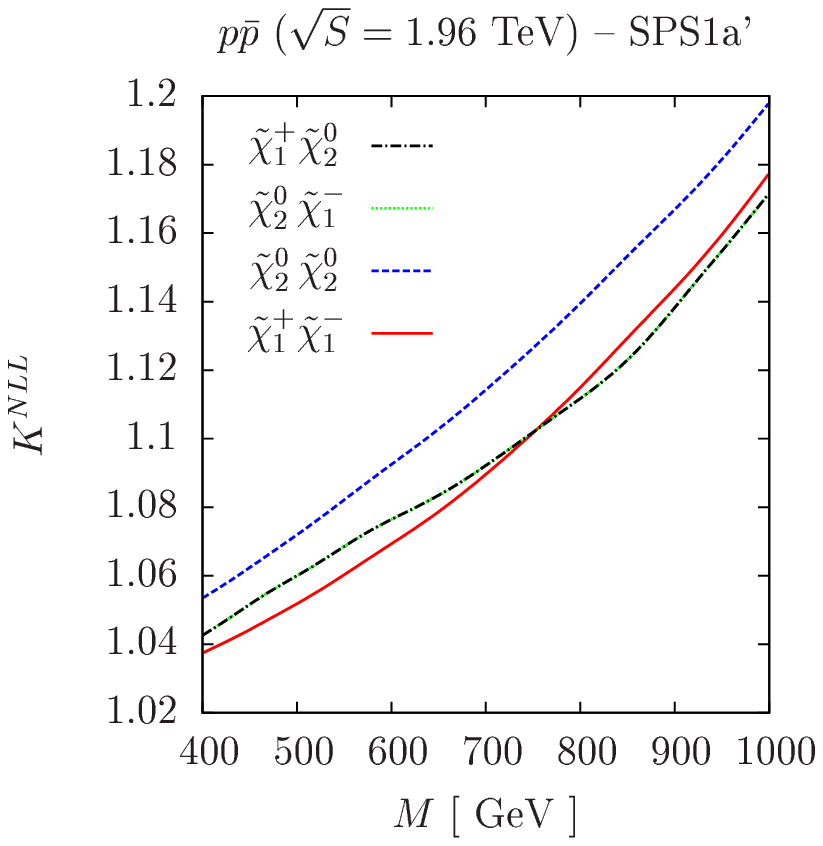,width=.9\textwidth}
	\caption{NLL $K$-factor, as defined in Eq.~\eqref{eq:3}, for the associated production of charginos and neutralinos (dotted and dot-dashed) as well as chargino (full) and neutralino (dashed) pairs at the Tevatron. \hfill
	\label{fig:2}}
    \end{minipage}
\end{figure}
In Fig.~\ref{fig:1}, we show the invariant mass spectrum of neutralino-chargino pairs at the Tevatron with centre-of-mass energy of $\sqrt{S}=1.96$ TeV.
Here, the $\cO(\alpha_s^0)$ cross section (LO) is evaluated with the leading-order parton distribution functions {\tt CTEQ6L1}.
While the $\cO(\alpha_s)$ corrections (NLO) are large and positive, they do not clearly improve the scale uncertainties obtained by varying the factorisation and renormalisation scales ($\mu$) in the range $[M/2, 2M]$.
However once they are matched to the resummed cross section at next-to-leading logarithmic accuracy, the resulting prediction (NLL+NLO) is found to be very stable and precise.

The impact of the resummation of the threshold-enhanced terms is shown for several gaugino pairs in Fig.~\ref{fig:2}.
Setting $\mu=M$ and defining the NLL $K$-factor by
\begin{equation}
    K^{NLL} = \frac{d \sigma^{\rm NLL+NLO}}{d \sigma^{\rm NLO}},
    \label{eq:3}
\end{equation}
we see that the NLL contributions increase the NLO predictions by a few percent in the low-$M$ region and can reach 20 \% for large invariant masses.
Note that threshold resummation has already been applied to the associated production of neutralino and chargino in Ref.~\cite{Li:2007ih}.
A careful comparison between their results and ours can be found in Ref.~\cite{Debove:2010kf}.

\begin{figure}[t]
    \begin{minipage}[t]{.48\columnwidth}
	\epsfig{figure=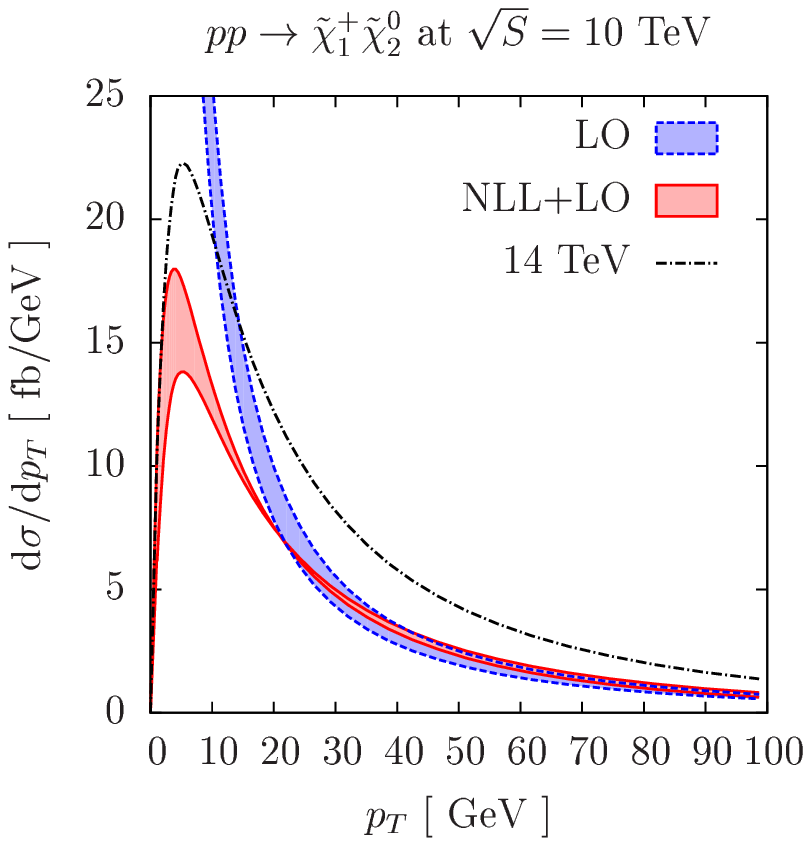,width=.9\textwidth}
	\caption{Transverse-momentum distributions of chargino-neutralino pairs at the LHC.
	The LO calculation (dashed) is matched to the resummed calculation (full).
	The scale uncertainty is shown as a shaded band and the matched result for the LHC design energy of 14 TeV as a dot-dashed line. \hfill
	\label{fig:3}}
    \end{minipage} \hfill
    \begin{minipage}[t]{.48\columnwidth}
	\epsfig{figure=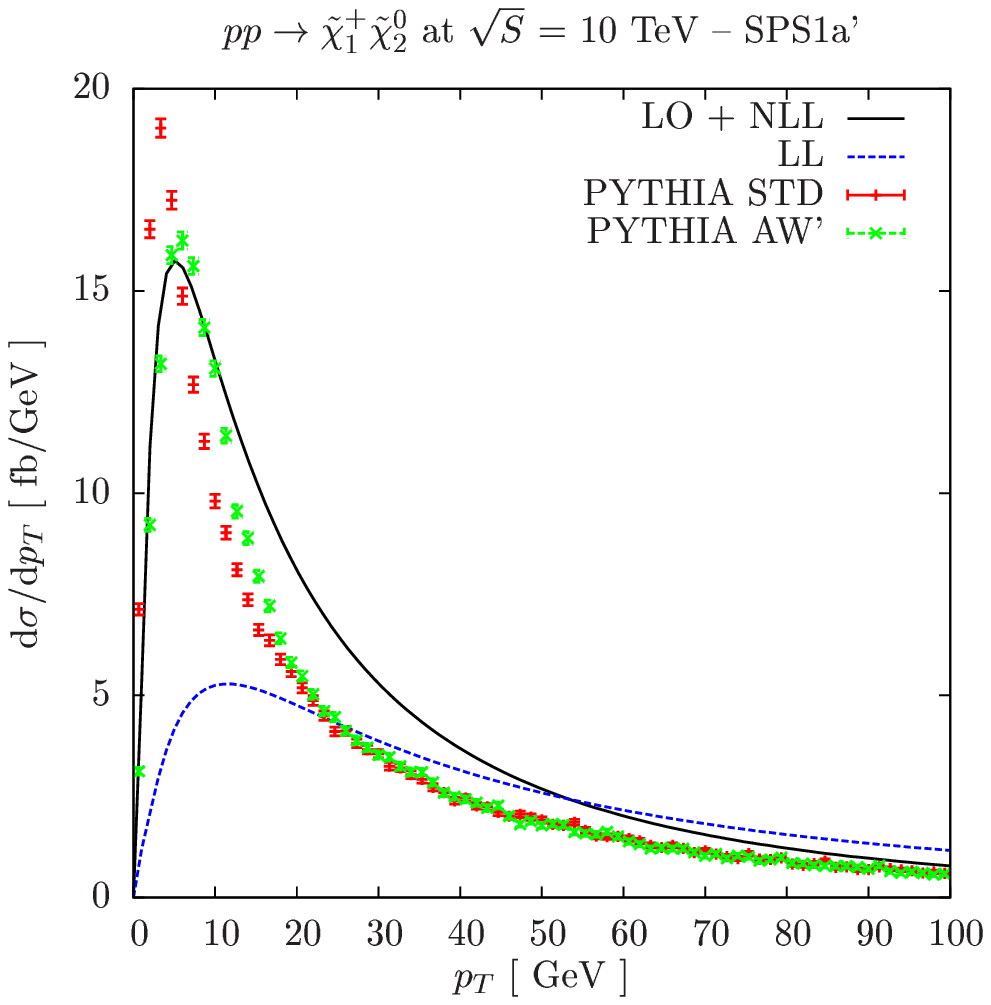,width=.85\textwidth}
	\caption{Transverse-momentum distributions of chargino-neutralino pairs at the LHC.
	The matched LO+NLL (full) and the LL (dashed) results are compared with the predictions of the PYTHIA parton shower with default (bars) and tuned (crosses) parameters. \hfill
	\label{fig:4}}
    \end{minipage}
\end{figure}
In Fig.~\ref{fig:3}, we show the $p_T$-spectrum of chargino-neutralino pairs produced at the LHC with a hadronic centre-of-mass energy of $\sqrt{S}=10$ GeV. 
The $\cO(\alpha_s)$ calculation (LO) diverges at low $p_T$ due to the logarithmic terms of the form $\alpha_s \ln(M^2/p_T^2)/p_T^2$, thus being totally unreliable. 
However, it becomes finite after having been matched to the resummed prediction at next-to-leading logarithmic (NLL+LO) accuracy. 
We also evaluate the uncertainties coming from scale variations in the range $[m_{\tilde{\chi}}/2, 2m_{\tilde{\chi}}]$.
The scale dependence of the NLL+LO prediction is clearly improved with respect to the LO result.
For comparison, the NLL+LO prediction for the 14 TeV design energy of the LHC is also presented.

In Fig.~\ref{fig:4}, we compare our NLL+LO prediction with our resummed result at leading-logarithmic (LL) accuracy and two different setups for the \texttt{PYTHIA6.4} \cite{Sjostrand:2006za} Monte Carlo (MC) generator.
We see that the default (STD) MC simulation is clearly improved beyond the LL approximation and approaches our NLL+LO result, but peaks at slightly smaller values of $p_T$.
This behaviour can be improved by tuning the intrinsic $p_T$ of the partons in the hadron (AW'),\cite{Debove:2009ia} but both simulations underestimate the intermediate $p_T$-region.
Results for the transverse-momentum distributions of neutralino and chargino pairs can be foud in Ref.~\cite{Debove:2009ia}.

\section{Conclusion}
In this work, threshold and transverse-momentum resummations have been presented for gaugino pair hadroproduction.
We found that the NLL contributions are important especially for the $p_T$-distributions which are not even finite using standard perturbation theory.
Our resummed predictions show a better stability under unphysical scale variations than the fixed-order calculations.
Finally we have demonstrated that our results modify considerably the commonly used MC predictions.
All these features will possibly lead to improvements for the experimental determination of the gaugino parameters.

\section*{Acknowledgements}
We thank M.~Klasen for useful discussions and comments.
\section*{References}
\providecommand{\href}[2]{#2}\begingroup\raggedright\endgroup
\end{document}